\documentclass[12pt]{article} 
\usepackage{amssymb}

\newcommand{\nc}{\newcommand}
\nc{\la}{\lambda} \nc{\alf}{\alpha} \nc{\La}{\Lambda} \nc{\ze}{\zeta}
\nc{\tht}{\theta} \nc{\T}{\Theta} \nc{\be}{\beta}  \nc{\eps}{\epsilon} 
\nc{\ga}{\gamma}  \nc{\De}{\Delta}  \nc{\G}{\Gamma}  \nc{\vphi}{\varphi}
\nc{\de}{\delta} \nc{\si}{\sigma}  \nc{\ka}{\kappa}   \nc{\Si}{\Sigma} 
\nc{\om}{\omega}  \nc{\qq}{\quad\quad}                \nc{\Om}{\Omega}
\nc{\nf}{\infty}   \nc{\dl}{\mathop{\smash{\cal L}}}  \nc{\black}{\rule{3mm}{3mm}}
\nc{\ra}{\rightarrow}    \nc{\ol}{\overline}        \nc{\und}{\underline} 
\nc{\beq}{\begin{equation}}  \nc{\eeq}{\end{equation}}  \nc{\pt}{\partial}  
   \nc{\dst}{\displaystyle}  \nc{\na}{\nabla} 
\nc{\nnb}{\nonumber}    \nc{\bs}{\backslash}        \nc{\mb}{\mathbb}   
\nc{\sn}{{\rm sn}\,} \nc{\cn}{{\rm cn}\,}     \nc{\dn}{{\rm dn}\,} \nc{\nin}{\noindent}
\nc{\ti}{\tilde}   \nc{\wti}{\widetilde}   \nc{\h}{\hat}  \nc{\wh}{\widehat}
\nc{\tpsi}{\wti{\psi}}   \nc{\tphi}{\wti{\phi}}  \nc{\tH}{\wti{H}} \nc{\Ai}{{\rm Ai}}

\newcounter{muni}
\newenvironment{remunerate}{\begin{list}{{\rm \arabic{muni}.}}
{\usecounter{muni}
\setlength{\leftmargin}{0pt}\setlength{\itemindent}{38pt}}}{\end{list}}

\nc{\brm}{\begin{remunerate}}   \nc{\erm}{\end{remunerate}}

\newtheorem{nth}{Proposition}
\nc{\stg}{\mathop{\smash{*}}}
\nc{\st}{\mathop{\smash{\delta}}}
\nc{\barr}{\begin{array}}   \nc{\earr}{\end{array}}   \nc{\dg}{\dagger}
\nc{\mtvb}{\mathversion{bold}}   \nc{\mtvn}{\mathversion{normal}}

\begin{document} 

\begin{titlepage}

\hfill{22 february 2011}

\vspace{1cm}
\centerline{\Large\bf  Further results on non-diagonal }

\vspace{4mm}
\centerline{\Large\bf Bianchi type III vacuum metrics }

\vskip 2.0truecm
\centerline{\large\bf Galliano VALENT}

\vskip 2.0truecm
\centerline{ \it Laboratoire de Physique Th\'eorique et des
Hautes Energies}
\centerline{\it CNRS, UMR 7589}
\centerline{\it 2 Place Jussieu, F-75251 Paris Cedex 05, France}
\nopagebreak

\vskip 2.5truecm

\begin{abstract}
We present the derivation, for these vacuum metrics, of the Painlev\'e VI equation first 
obtained by Christodoulakis and Terzis, from the field equations for both minkowskian and 
euclidean signatures. This allows a complete discussion and the precise connection with some 
old results due to Kinnersley. The hyperk\"ahler metrics are shown to belong to the Multi-Centre 
class and for the cases exhibiting an integrable geodesic flow the relevant Killing tensors 
are given. We conclude by the proof that for the Bianchi B family, excluding  type III, there are 
no hyperk\"ahler metrics.
\end{abstract}

\end{titlepage}

\section{Introduction}
The study of exact solutions for empty space Einstein equations with Bianchi type B isometries 
has been worked out by Christodoulakis and Terzis. In a series of groundbreaking papers \cite{ctbis}, \cite{ct}, \cite{tc1}, \cite{tc2} these authors have obtained the most general form of the mixing 
matrix and reduced the field equations to Painlev\'e VI for the types: III,$\ $IV and VII$_h$, the 
more difficult type VI$_h$ remaining still unsolved. 

The special case of Bianchi type III empty space metrics was analyzed in \cite{ctbis} and \cite{ct}. 
Despite the many results obtained in these two references, some questions were left aside :
\brm
\item Their differential system, given by their equations (2.8) to (2.10) in reference \cite{ct}, does solve Einstein equations. However since it was not derived from the field equations it is not 
equivalent to them and the analysis cannot be claimed to be fully general.  
\item A new interesting euclidean metric was found while the authors were looking for minkowskian metrics. Why is it so?
\item They found several explicit minkowskian metrics of Petrov type D. What is their precise relation with Kinnersley metrics  \cite{Ki}? 
\item Among all of the euclidean metrics which ones are hyperk\"ahler?
\erm
It is the aim of this article to clarify these points and to study which metrics, within this family, exhibit an integrable geodesic flow.

In Section 2 we write down the field equations for both minkowskian and euclidean signatures in 
which a parameter $\ka$ play a prominent role. In Section 3 the special case $\,\ka=0$ is first considered  
leading to Kinnersley metrics and their euclidean partners. It is then possible to relate 
precisely all of the Petrov type D metrics found in \cite{ct} with Kinnersley's metrics. Among 
the euclidean metrics there is a ``little" hyperk\"ahler metric which is shown to belong to the Multi-Centre family. In Section 4 the general case, for which $\,\ka\neq 0$, is discussed and 
it is shown that all the functions appearing in the metric can be expressed in terms of a 
single function $\,\mu(t)$ and its derivatives. It follows in Section 5 that $\mu(t)$ satisfies 
a non-linear second order ordinary differential equation, which can be related, using some 
results due to Okamoto, to Painlev\'e VI and which is different from the one found in \cite{ct}. 
Using a classical solution we retrieve the euclidean metric first discovered in \cite{ct}, 
which we show to be hyperk\"ahler and it reduces, in some limit, to the ``little" hyperk\"ahler 
metric. In Section 6 the case of a minkowskian signature singles out a parameter $E$, 
and only for $\,E\neq 0$ do we get Painlev\'e VI, while for the special value $E=0$ a 
Lie symmetry allows for integration and gives Siklos metric. In Section 7 the metrics 
with an integrable geodesic flow are determined and their (quadratic) Killing-St\"ackel tensor 
is constructed. Eventually, in Section 8, it is proved that for the Bianchi B family, except for type III, 
there can be no hyperk\"ahler metric.

\section{Derivation of the field equations}
For the Bianchi type III Lie algebra the Maurer-Cartan 1-forms are:
\[\si_1=dx,\qq\si_2=dy,\qq\si_3=e^{-x}\,dz\quad\Rightarrow\quad d\si_1=d\si_2=0,
\qq d\si_3=\si_3\wedge\si_1,\]
with the  Killing fields and non-vanishing commutator
\[{\cal L}_1=\pt_x+z\,\pt_z,\qq {\cal L}_2=\pt_y,\qq{\cal L}_3=\pt_z\qq\quad  
[{\cal L}_3,{\cal L}_1]={\cal L}_3.\]

As shown in \cite{ct} the most general non-diagonal metric allows for a mixing of the $\si_2$ and 
$\si_3$ forms. We found convenient to  write the metric
\beq\label{classemet}
g=\eps\,\alf^2\,dt^2+\be^2\,\si_1^2+\ga^2\,\si_3^2+\de^2(\si_2+\mu\,\si_3)^2,\qq\eps=\pm 1,\eeq
where all the functions involved depend solely on $t$.

We will follow Geroch analysis \cite{Ge} with respect to the Killing vector ${\cal L}_2=\pt_y$ 
and write the metric
\beq
g=\frac 1V(dy+\Theta)^2+V\,\Gamma,\eeq
with
\beq\left\{\barr{l}\dst 
V=\frac 1{\de^2}\qq\Theta=\mu\,e^{-x}\,dz\\[4mm]
\G=\eps\,\alf^2\de^2\,dt^2+\be^2\de^2\,dx^2+\ga^2\de^2\,\si_3^2=e_1^2+e_2^2+e_3^2.\earr\right.\eeq

\nin The empty space Einstein equations are equivalent to 
\beq\label{eqE0}\barr{ll}
(a)\qq & \dst *_{\G}\,\frac{d\Theta}{V^2}=d\Psi,\\[4mm] 
(b)\qq & \dst
ric_{ij}(\G)=\frac{V^2}{4}(\pt_i\chi_+\,\pt_j\chi_-+\pt_j\chi_+\,\pt_i\chi_-),\\[4mm]
(c)\qq & \dst \frac{\Delta V}{V}=-2\,r(\G),\earr \qq \chi_{\pm}=\frac 1V\pm\Psi,\eeq
where $\,ric_{ij}(\G)$ and $\,r(\G)$ are the Ricci tensor and the scalar curvature of the metric $\G$. 
An easy computation gives
\[d\Psi=*_{\G}\,\frac{d\Theta}{V^2}=-\frac{\dot{\mu}\be\de^3}{\alf\ga}\,dx
-\frac{\mu\alf\de^3}{\be\ga}\,dt\]
and its integrability condition is nothing but
\beq\label{eqmu}
\dot{\mu}=\ka\,\frac{\alf\ga}{\be\de^3}\quad\Rightarrow\quad \pt_x\Psi=
-\ka\qq\pt_t\Psi=-\frac{\mu\alf\de^3}{\be\ga}\eeq
where $\ka$ is some real constant. 

The non-diagonal equation for $ric_{ij}(\G)$ gives a first order relation, which we collect with (\ref{eqmu}):
\beq\label{eqspe}
\frac{\dot{\be}}{\be}=\frac{\dot{\ga}}{\ga} +\frac{\ka}{2}\frac{\alf\mu}{\be\ga\de},
\qq\qq \dot{\mu}=\ka\,\frac{\alf\ga}{\be\de^3}.\eeq
The remaining ones, after taking some combinations, are most conveniently written
\beq\label{eqE}\barr{ll} 
(a)\qq & \dst\frac{\ddot{\be}}{\be}
+\frac{\dot{\be}}{\be}\left(\frac{\dot{\ga}}{\ga}+\frac{\dot{\de}}{\de}-\frac{\dot{\alf}}{\alf}\right)
+\eps\,\frac{\alf^2}{\be^2}+\frac{\eps}{2}\frac{\alf^2\de^2\mu^2}{\be^2\ga^2}=0\\[4mm]
(b)\qq & \dst \frac{\ddot{(\ga\de)}}{\ga\de}
+\frac{\dot{(\ga\de)}}{\ga\de}\left(\frac{\dot{\be}}{\be}-\frac{\dot{\alf}}{\alf}\right)
+\eps\,\frac{\alf^2}{\be^2}=0\\[4mm]
(c)\qq & \dst 2\left(\frac{\dot{\be}\dot{\ga}}{\be\ga}+\frac{\dot{\ga}\dot{\de}}{\ga\de}
+\frac{\dot{\de}\dot{\be}}{\de\be}\right)+2\eps\,\frac{\alf^2}{\be^2}
+\frac{\eps}{2}\,\frac{\alf^2\de^2\mu^2}{\be^2\ga^2}-\frac{\ka^2}{2}\frac{\alf^2}{\be^2\de^4}=0\earr\eeq
Let us notice that the differential equation for $\,\ga$, which is
\beq\label{eqGa} 
\frac{\ddot{\ga}}{\ga}
+\frac{\dot{\ga}}{\ga}\left(\frac{\dot{\be}}{\be}+\frac{\dot{\de}}{\de}-\frac{\dot{\alf}}{\alf}\right)
+\eps\,\frac{\alf^2}{\be^2}+\frac{\eps}{2}\frac{\alf^2\de^2\mu^2}{\be^2\ga^2}+\frac{\ka^2}{2}\frac{\alf^2}{\be^2\,\de^4}=0\eeq
follows from (\ref{eqE}(a)) and the first relation in (\ref{eqspe}).
 
\section{The special case $\ka=0$}
Since this case should correspond to Kinnersley metric let us derive it shortly. The function  $\mu=\mu_0$ is constant, and (\ref{eqmu}) implies $\be=c\,\ga$ for some constant $c\neq 0$. Taking $\alf=1/\de$ to fix up the time coordinate, equation (\ref{eqE}(b)) gives
\[\ga\de\ddot{(\ga\de)}+\dot{(\ga\de)}^2+\frac{\eps}{c^2}=0\quad\Rightarrow\quad (\ga\de)^2=a-2mt-\frac{\eps}{c^2}\, t^2,\]
where $a,m$ are integration constants. The equation (\ref{eqGa}) becomes
\[D_t\left(\frac{\dot{\ga}}{\ga}\right)+2\frac{\dot{\ga}^2}{\ga^2}
+2\frac{\dot{\ga}}{\ga}\frac{\dot{\de}}{\de}+\frac{\eps\mu_0^2}{2}\frac 1{c^2\ga^4}
+\frac{\eps}{c^2\ga^2\de^2}=0\]
and relation (\ref{eqE}(c)) 
\beq\label{cons1}
\frac{\dot{\ga}^2}{\ga^2}+2\frac{\dot{\ga}}{\ga}\frac{\dot{\de}}{\de}
+\frac{\eps\mu_0^2}{4}\frac 1{c^2\ga^4}+\frac{\eps}{c^2\ga^2\de^2}=0.\eeq
Subtracting these relations we obtain 
\beq
\ddot{\ga}+\frac{\eps\mu_0^2}{4}\frac 1{\ga^3}=0\quad\Rightarrow\quad \dot{\ga}^2=E+\frac{\eps\mu_0^2}{4c^2}\frac 1{\ga^2}.
\eeq

\subsection{Minkowskian signature and Kinnersley metric}
For $\eps=-1$ we must have $E>0$. One gets 
\[\ga^2=E\left(t^2+\frac{\mu_0^2}{4c^2E^2}\right)\] 
and imposing (\ref{cons1}) we have
\[c^2=1\qq\qq a=-\frac{\mu_0^2}{4E^2}.\]
So it is convenient to define
\[l=\frac{\mu_0}{2E}\quad\Rightarrow\quad \ga^2=E(t^2+l^2)\]
and, up to scalings, we get the metric
\beq\label{Ki}
\frac{g}{E}=-\frac{dt^2}{U}+(t^2+l^2)(\si_1^2+\si_3^2)+U(\si_2+2l\,\si_3)^2
\qq U=\frac{t^2-2mt-l^2}{t^2+l^2}\eeq
which is Petrov type D. Since {\em all} the type D vacuum metrics were classified by Kinnersley, 
let us give the precise relation to his work. 

Let us consider specifically \cite{Ki} his II-D metric. Up to slight notational changes one has 
\beq\label{k1}\barr{l}\dst 
g=-\frac{\rho}{\De}dt^2
+\frac{\De}{\rho}\left(du+(2l\,e^{-x}+a\,e^{-2x})dv-\frac{\rho}{\De}dt\right)^2\\[4mm]\dst 
\hspace{7cm}+\rho\,dx^2+\frac{e^{-2x}}{\rho}\Big(adu-(t^2+l^2)dv\Big)^2.\earr\eeq
with
\[\De=t^2-2mt-l^2,\qq\quad \rho=t^2+(l+a\,e^{-x})^2.\]
The transformations
\[dy=du-\frac{(t^2+l^2)}{\De}dt,\qq dz=dv-\frac a{\De}dt,\]
bring (\ref{k1}) to a Boyer-Lindquist form
\beq\label{k2}
g=-\frac{\rho}{\De}dt^2+\frac{\De}{\rho}(dy+(2l\,e^{-x}+a\,e^{-2x})dz)^2+\rho\,dx^2
+\frac{e^{-2x}}{\rho}(ady-(t^2+l^2)dz)^2.\eeq
This type D metric has only two commuting Killing vectors $\pt_y$ and $\pt_z$ but, as observed 
by \cite{Am}, in the limit $a\to 0$, there appear 4 Killing vectors, and the metric (\ref{k2}) 
does transform into (\ref{Ki}). The Bianchi type III isometries are manifest and an extra fourth 
Killing vector 
\beq\label{L4}
{\cal L}_4=z\,\pt_x+2l\,e^x\,\pt_y+\frac 12(z^2-e^{2x})\pt_z\eeq
leaves the metric invariant because of the relations
\[{\cal L}_4\,\si_2=2l\,e^x\,\si_1\qq \qq  
{\cal L}_4\,\si_3=-e^x\,\si_1 \qq\qq {\cal L}_4\,\si_1=+e^x\,\si_3.\]
These isometries have for non-vanishing commutators 
\beq 
[{\cal L}_3,{\cal L}_1]={\cal L}_3,\qq [{\cal L}_1,{\cal L}_4]={\cal L}_4\qq 
[{\cal L}_3,{\cal L}_4]={\cal L}_1.\eeq
So we do realize that Kinnersley had discovered in 1969 the first non-diagonal Bianchi 
type III metric! 

The limit  $l\to 0$ allows, by a scaling of $t$, to take $2m=1$ and gives the diagonal metric
\beq
g_0=-\frac{dt^2}{1-1/t}+t^2(\si_1^2+\si_3^2)+(1-\frac 1t)\si_2^2,\eeq
which was derived just a year before by Stewart and Ellis \cite{se} and by Cahen and Defrise 
\cite{cd} and re-discovered later in \cite{mt} and \cite{mm}.

As observed by Christodoulakis and Terzis some of their metrics \cite{ct} being Petrov type D 
should be special cases of Kinnersley. A first example is given in their formula (2.37). 
Using the relation $B=C+4A$ and up to a change of the normalization of $\si_3$ we get
\beq
g_1=-A\,d\xi^2+A(\si_1^2+\si_3^2)+C(\si_2+\frac 12\,\si_3)^2,\eeq
with
\[A=\frac 12(\cosh^2 \xi+2\la\cosh\xi+1),\qq C=\frac{4(1-\la^2)\sinh^2 \xi}{A},\qq \la\in\,]-1,+1[.\]
The change of variable and of parameter 
\[\cosh \xi=\frac{t-m}{\sqrt{m^2+l^2}},\qq \la=\frac m{\sqrt{m^2+l^2}},\qq l=\frac 14\]
shows that $g_1$ is homothetic to Kinnersley metric (\ref{Ki}) for the special 
value of $\,l\,$ given above.

A second example is their metric:
\beq
g_2=-\frac{e^{2\xi}(e^{2\xi}+1)}{4(2e^{2\xi}+1)}\,d\xi^2+\frac{e^{\xi}\cosh\xi}{4}(\si_1^2+\si_3^2)
+\frac{e^{\xi}}{2\cosh\xi}(\si_2+\frac 12\,\si_3)^2.\eeq
The change of variable and the choice of parameters
\[e^{\xi}=2\sqrt{2(t^2-1/16)},\qq m=0,\qq l=\frac 14,\]
show that $g_2$ is again Kinnersley metric (\ref{Ki}) for the special values of $\,(m,l)\,$ given above. 
Let us consider now the euclidean signature.

\subsection{Euclidean signature}
For $\eps=+1$, according to the sign of $E$, the metrics are:
\beq\label{kazero}\barr{lll}
E>0 & \dst g_+=\frac{dt^2}{U_+}+(t^2-l^2)(\si_1^2+\si_3^2)+U_+(\si_2 +2l\si_3)^2
\quad & \dst U_+=-\frac{t^2-2mt+l^2}{t^2-l^2}\\[4mm]
E=0 & \dst g_0=\frac{dt^2}{U_0}+t(\si_1^2+\si_3^2)+U_0(\si_2+\si_3)^2
\quad & \dst U_0=\frac{a^2-t^2}{t}\\[4mm]
E<0 & \dst g_-=\frac{dt^2}{U_-}+(l^2-t^2)(\si_1^2+\si_3^2)+U_-(\si_2+2l\si_3)^2
\quad & \dst U_-=-\frac{t^2-2mt+l^2}{l^2-t^2} \earr\eeq
They all exhibit a fourth Killing vector given by (\ref{L4}). Let us also notice that despite 
the relation $g_+=-g_-$ the two metrics for $\,E\neq 0$ will be different  
since positivity will give definitely different intervals of variation for $t$.

We will call the metric for $E=0$ the ``little" hyperk\"ahler metric since it is a special case 
of the more general hyperk\"ahler metric that will be discussed later in Section 5.

Let us prove :

\begin{nth}
The metric in (\ref{kazero}) for $\,E=0$ is hyperk\"ahler (or with self-dual curvature) since 
it belongs to the Multi-Centre family.
\end{nth}

\nin{\bf Proof:}
The reader may consult \cite{vy} for the definition and some basic properties of the 
Multi-Centre metrics. Using for preferred Killing $\,\pt_z$ the metric becomes 
\beq
g=\frac 1V(dz+\T)^2+V\,\G,\eeq
with
\beq
V=\frac t{a^2X^2}\qq\T=\frac{a^2-t^2}{a^2X}\,dy\qq X=e^{-x}\eeq
and the three dimensional metric
\beq
\G=a^2\left\{dX^2+\frac{X^2}{(a^2-t^2)}dt^2+X^2\frac{(a^2-t^2)}{a^2}dy^2\right\}=
du^2+dv^2+dw^2\eeq
with the flattening coordinates 
\[u=X\sqrt{a^2-t^2}\cos y\qq v=X\sqrt{a^2-t^2}\sin y\qq w=X\,t.\]
The potential and the gauge field become
\beq
V=a\,\frac{w}{r^3}\qq\qq \T=-a\,\frac{(v\,du-u\,dv)}{r^3}\qq\quad r=\sqrt{u^2+v^2+w^2}.\eeq
The potential is that of a dipole located at the origin and aligned with the $w$ axis. It  
obviously satisfies $\,\De\,V=0$ proving that the metric is indeed a 
Multi-Centre, hence hyperk\"ahler. 
The triplet of covariantly constant complex structures $\,\Om_i$ is
\beq\label{complex}\left\{\barr{l}
\Om_1=(dz+\T)\wedge du+V\,dv\wedge dw\\[4mm]
\Om_2=(dz+\T)\wedge dv+V\,dw\wedge du\\[4mm]
\Om_3=(dz+\T)\wedge dw+V\,du\wedge dv,\earr\right.\eeq
and this ends the proof. $\quad\Box$

The Killing vectors, in these coordinates, become
\[\barr{l}
{\cal L}_1=z\,\pt_z-u\,\pt_u-v\,\pt_v-w\,\pt_w\qq{\cal L}_2=-v\,\pt_u+u\,\pt_v\qq {\cal L}_3=\pt_z\\[4mm]
\dst {\cal L}_4=\frac 12\left(z^2-\frac{a^2}{r^2}\right)\pt_z-\left(zu+a\frac vr\right)\pt_u
-\left(zv-a\frac ur\right)\pt_v-zw\,\pt_w.\earr\]
It follows that $\,{\cal L}_1,\,{\cal L}_3,\,{\cal L}_4$ are tri-holomorphic 
\[{\cal L}_a\,\Om_i=0\qq a=1,3,4\qq i=1,2,3\]
while $\,{\cal L}_2$ is holomorphic  
\[{\cal L}_2\,\Om_1=-\Om_2\qq {\cal L}_2\,\Om_2=\Om_1\qq {\cal L}_2\,\Om_3=0.\] 
Let us consider now the general case for which $\ka\neq 0$.

\section{General case}
A convenient choice of time coordinate is  
$\ \ka\,\alf=\be\,\ga\,\de$ which simplifies the two relations in (\ref{eqspe}) to
\beq\label{muetdmu}
\dot{\mu}=\frac{\ga^2}{\de^2}>0\qq\qq \frac{\mu}{2}=\frac{\dot{\be}}{\be}-\frac{\dot{\ga}}{\ga}
\eeq
and allows to express all functions in terms of $\,\ga\de,\,\mu$ and their derivatives:
\beq\label{fctsmu}
\ga^2=\ga\de\,\sqrt{\dot{\mu}}\qq\de^2=\frac{\ga\de}{\sqrt{\dot{\mu}}}\qq 
2\,\frac{\dot{\be}}{\be}=\frac{\ddot{\mu}}{2\dot{\mu}}+\mu+\frac{\dot{(\ga\de)}}{\ga\de}
\qq \alf^2=\be^2\,\frac{(\ga\de)^2}{\ka^2}.\eeq
By a scaling on $\ga$ and $\de$ we can set $\ka=1$ and equation (\ref{eqE}(b)) which becomes
\beq
D_t\left(\frac{\dot{(\ga\de)}}{\ga\de}\right)+\eps\,(\ga\de)^2=0\eeq
can be integrated once. So we are left with the differential system: 
\beq\label{sysODE}\barr{ll}
(a)\qq & \dst 
\left(\frac{\dot{(\ga\de)}}{\ga\de}\right)^2+\eps\,(\ga\de)^2=E\\[4mm]
(b)\qq & \dst 
\left(\frac{\ddot{\mu}}{2\dot{\mu}}-\frac{\dot{(\ga\de)}}{\ga\de}\right)^2+\dot{\mu}
-2\frac{\dot{(\ga\de)}}{\ga\de}\,\mu-4E-\eps\,(\ga\de)^2\,\frac{\mu^2}{\dot{\mu}}=0
\earr\qq\dot{\mu}>0\eeq
where $\,E$ is some real constant. Let us notice that equation (\ref{eqE}(a)), formerly the 
differential equation for $\,\be$, becomes 
\beq\label{odebeta}
D_t\,A+\dot{\mu}+\eps\,(\ga\de)^2\,\frac{\mu^2}{\dot{\mu}}=0
\qq\qq A=\frac{\ddot{\mu}}{2\dot{\mu}}-\frac{\dot{(\ga\de)}}{\ga\de}\eeq
and should appear as equation (\ref{sysODE}(c)). However differentiating (\ref{sysODE}(b)) we get
\beq\label{test}
A\left(D_t\,A+\dot{\mu}+\eps\,(\ga\de)^2\,\frac{\mu^2}{\dot{\mu}}\right)=0.\eeq
Since the relations in (\ref{odebeta}) imply that $\,A$ cannot vanish identically, 
the equation (\ref{odebeta}) is a consequence of (\ref{sysODE}(b)), showing that this last equation is an integrated form of (\ref{odebeta}). 

Let us first consider the euclidean signature.

\section{Euclidean signature}
In this case $\,E>0$, and the general solution of (\ref{sysODE}(a)) can be written
\beq
\ga\de=\frac{\sqrt{E}}{\cosh(\sqrt{E}t)}.\eeq
By the scalings $\mu\to \mu/\sqrt{E}$ and $t\to \sqrt{E}\,t$ we can take $\,E=1$. The  
ode (\ref{sysODE}(b)) becomes
\beq\label{ode2}
\left(\frac{\ddot{\mu}}{2\dot{\mu}}+\tanh t\right)^2+\dot{\mu}+2\tanh t\,\mu-4
-(1-\tanh^2 t)\,\frac{\mu^2}{\dot{\mu}}=0 \qq \dot{\mu}>0\eeq
and the metric
\beq
g=\be^2\left(\frac{dt^2}{\cosh^2 t}+\si_1^2\right)+\frac{\sqrt{\dot{\mu}}}{\cosh t}\,\si_3^2
+\frac 1{\cosh t\,\sqrt{\dot{\mu}}}\,(\si_2+\mu\,\si_3)^2\eeq
with
\beq
\frac{\dot{(\be^2)}}{\be^2}=\frac{\ddot{\mu}}{2\dot{\mu}}+\mu-\tanh t.\eeq
It is convenient to make the transformations
\[x=(1+\tanh t)/2\qq\qq  \mu(t)=-\frac 12\,m(x)\qq\qq m'=D_x\,m\]
which bring the ode (\ref{ode2}) into
\beq\label{ode3}
x^2(1-x)^2(m'')^2=-4m'(xm'-m)^2+4m'^2(xm'-m)+4m'^2,\qq x\in\,(0,1).\eeq 

This ODE is clearly different from the one obtained by Christodoulakis and Terzis in \cite{cs}. 
In their work they used results of Cosgrove and Scoufis \cite{cs} to reduce (\ref{ode3}) 
to Painlev\'e VI. We will not follow this path since Okamoto \cite{Ok} gave much simpler 
B\"acklund transformations. Using Okamoto notations let us take  
$\,b_2=0$ and denote $\,{\bf b}=(b_1,b_3,b_4)$. Let us consider the 
solution $y(x,{\bf b})$ of Painlev\'e VI with parameters
\[\alf=\frac 12\,(b_3-b_4)^2\qq\be=-\frac 12\,b_1^2\qq\ga=\frac 12\,b_1^2\qq
\de=\frac 12\big[ 1-(b_3+b_4+1)^2\big].\]
If one defines
\beq
h(x)=x(x-1)\,H(x)+s_2({\bf b})\left(x-\frac 12\right),\eeq
where $H$ is the Hamilton function
\beq\label{Hp6}\left\{\barr{l} 
x(x-1)\,H=y(y-1)(y-x)\,z^2-[b_1(2y-1)(y-x)+(b_3+b_4)y(y-1)]z\\[4mm]
\hspace{7cm}+(b_1+b_3)(b_1+b_4)(y-x)\\[4mm]
\dst 2z=\frac{x(x-1)y'}{y(y-1)(y-x)}+\frac{b_1\,(2y-1)}{y(y-1)}+\frac{b_3+b_4}{y-x},\earr
\right.\eeq
then $h$ is a solution of
\beq\label{Ok1}
x^2(x-1)^2(h'')^2=-4h'(xh'-h)^2+4(h')^2(xh'-h)+s_1({\bf b}^2)(h')^2+s_2({\bf b}^2)h'
+s_3({\bf b}^2)\eeq
where the $s_i({\bf b}^2)\ i=1,2,3$ are the symmetric polynomials for ${\bf b}^2=(b_1^2,\,b_3^2,\,b_4^2)$.
It follows that $\,m=h$ is a solution of (\ref{ode3}).

The identification of (\ref{Ok1}) and (\ref{ode3}) gives 3 different cases:
\beq\barr{lccc}
\quad\qq & (b_1,\,b_3,\,b_4) & \qq & (\alf,\,\be,\,\ga,\,\de)\\[4mm]
(a)\qq\qq & (0,+2,0)  & \qq & (2,0,0,-4)\\[4mm]
(b)\qq\qq & (0,-2,0) & \qq & (2,0,0,0)\\[4mm]
(c)\qq\qq & (2,0,0)  & \qq & (0,-2,2,0)\earr\qq\quad (b_2=0).\eeq
The B\"acklund transformation follows from (\ref{Hp6}); for the first two cases
\beq
(a)\quad (b)\quad\to\qq m=\frac 1{4y(y-1)(y-x)^2}\Big\{x^2(x-1)^2\,(y')^2-4y^2(y-1)^2\Big\}\eeq
while for the third case
\beq
(c)\quad\to\qq m=\frac 1{4y(y-1)(y-x)^2}\Big\{x^2(x-1)^2\,(y')^2-4(y-x)^2\Big\}\eeq
The derivative of $\,m$ is
\beq\label{bac1}\barr{l}\dst 
(a)\quad \to\qq\qq m'=-\frac 1{4y(y-1)}\Big[\frac{x(x-1)\,(y')^2+2y(y-1)}{(y-x)}\Big]^2\\[4mm]\dst 
(b)\quad \to\qq\qq m'=-\frac 1{4y(y-1)}\Big[\frac{x(x-1)\,(y')^2-2y(y-1)}{(y-x)}\Big]^2\\[4mm]\dst 
(c)\quad \to\qq\qq m'=-\frac m{y-x}\earr
\eeq
so there are are further restrictions on $\,y$ since $m'$ must be negative.

\vspace{2mm}\nin {\bf Remarks:}
\brm
\item Using the birational transformations given by Okamoto, all of these solutions can be 
reduced to $\,{\bf b}=(0,0,0)$ which means that $\,(\alf,\,\be,\,\ga,\,\de)=(0,0,0,0)$. Despite the great simplicity of the parameters the complete solution is not known.
\item In the first two cases there are ``classical solutions" given by
\[(a):\qq y=\frac{(1-x)^2}{(1-x)^2+Cx^2}\qq\qq (b):\qq y=\frac{x^2}{2x-1+C(1-x)^2},\] 
but their B\"acklund transform (\ref{bac1}) gives $m\equiv 0$.
\item In the last two cases there are also classical solutions given by \cite{gls}. Only 
for case (b) do we get the non-trivial result
\beq\label{Gromov}
\frac m2=\frac{2x-1+2x(1-x)F}{1+x(1-x)\,F^2}\qq\quad F=\ln\left(\frac x{1-x}\right)+2a,
\qq a\in{\mb R}.\eeq
Going back to the $t$ variable it is convenient to define
\beq\label{hk2}
U=(t+a)\,\tanh t-1\qq a\in{\mb R}\qq\Longrightarrow\qq \ddot{U}+\frac 2{\cosh^2 t}\,U=0,\eeq
which allows to write
\beq
\frac{\mu}{2}=\frac{\dot{U}}{\dot{U}^2+U^2/\cosh^2 t}\qq\qq
\frac{\dot{\mu}}{4}=\frac{U^2/\cosh^2 t}{(\dot{U}^2+U^2/\cosh^2 t)^2},\eeq
and upon integration of relations (\ref{fctsmu}) one gets (up to 
scalings) the final form of this euclidean metric 
\beq\label{hk1}
g=U\left(\frac{dt^2}{\cosh^2 t}+\si_1^2\right)+\frac{U}{\cosh^2 t}\,\si_2^2
+\frac 1U\Big(\si_3+\dot{U}\,\si_2\Big)^2.\eeq

Let us observe that in the limit $a\to +\nf$, 
dividing the full metric by $a$ and changing $\si_3\to \si_3/a$ one gets  
\[g_{\nf}=\frac{\tanh t}{\cosh^2 t}\,dt^2+\tanh t(\si_1^2+\si_3^2)
+\frac{\cosh^2 t}{\tanh t}(\si_2+\si_3)^2\]
and upon the change of coordinate $u=a\,\tanh t$ one recovers the metric (\ref{kazero}) for $\,E=0$ 
which was already proved to be hyperk\"ahler.

In fact the euclidean metric (\ref{hk1}) is nothing but Christodoulakis and Terzis 
metric \cite{ct} written as:
\beq\label{christo}
A\,d\xi^2+B\,\si_1^2+D\,\si_2^2+2C\,\si_2\,\si_3+\frac 1B\,\si_3^2\eeq
where all functions depend solely on $\xi$. Noticing the relation
\[\frac 1B-\frac{C^2}{D}=\frac 1{\cosh^2(\mu^2\,\cos^2 \xi)D}\]
and using the variable $\,t=\mu^2\,\cos^2 \xi$, some algebra and several scalings 
bring the metric (\ref{christo}) to the form (\ref{hk1}).

Let us prove:

\begin{nth} The metric (\ref{hk1}) is hyperk\"ahler since it belongs to the 
Multi-Centre family.\end{nth}

\nin{\bf Proof:}  
We will use the Killing vector $\pt_{z}$ to write (\ref{hk1}) in the 
Multi-Centre form
\beq
g=\frac 1V\,(dz+\T)^2+V\,\G\eeq
with
\beq
X=e^{-x}\qq\qq\frac 1V=\frac{X^2}{U}\qq\qq\T=\frac{\dot{U}}{X}\,dy\eeq
and the three dimensional metric 
\beq
\G=dX^2+\frac{X^2}{\cosh^2 t}(dt^2+dy^2)=du^2+dv^2+dw^2\eeq
with the flattening coordinates
\beq 
u=\frac{X}{\cosh t}\,\cos y\qq\qq v=\frac{X}{\cosh t}\,\sin y\qq\qq 
w=X\,\tanh t.\eeq
The potential becomes
\beq
V=a\,\frac w{r^3}-\frac 1{r^2}+\frac w{2r^3}\,\ln\frac{r+w}{r-w},\qq\qq r=\sqrt{u^2+v^2+w^2}.\eeq
We have checked that 
$\,\De\,V=0$ establishing that this metric is a Multi-Centre and hence is hyperk\"ahler. The  
complex structures are still given by (\ref{complex}). This ends the proof.$\quad\Box$
\erm

Let us consider now the minkowskian signature.

\section{Minkowskian signature}
Now the final integration of equation (\ref{sysODE}(a)) does depend on the sign of $E$. We 
must analyze separately each case.

\subsection{First case: $\,\mathbf{E>0}$}
We have
\beq
\frac{\ga\de}{\ka}=\frac{\sqrt{E}}{\sinh(\sqrt{E}t)}.\eeq
Up to scalings of $\,\mu$ and $\,t$ we can take $\,E=1$. The ode (\ref{sysODE}(b)) becomes
\beq\label{finEp}
\left(\frac{\ddot{\mu}}{2\dot{\mu}}+\coth t\right)^2+\dot{\mu}+2\coth t\,\mu-4
+(\coth^2 t-1)\frac{\mu^2}{\dot{\mu}}=0\eeq
and the metric
\beq
g=\be^2\left(\frac{dt^2}{\sinh^2 t}+\si_1^2\right)+\frac{\sqrt{\dot{\mu}}}{\sinh t}\,\si_3^2
+\frac 1{\sinh t\,\sqrt{\dot{\mu}}}\,(\si_2+\mu\,\si_3)^2\eeq
with
\beq
\frac{\dot{(\be^2)}}{\be^2}=\frac{\ddot{\mu}}{2\dot{\mu}}+\mu-\coth t.\eeq
It is convenient to make the changes
\[x=\frac 12\,(\coth t+1) \qq\qq \mu(t)=-2\,m(x)\qq\qq  m'=D_x\,m.\]
The equation (\ref{finEp}) becomes 
\beq
x^2(x-1)^2\,(m'')^2=-4m'(xm'-m)^2+4m'^2(xm'-m)+4m'^2,\qq x\in(1,+\nf).\eeq
This ode is {\em exactly the same as for the euclidean signature} (compare with (\ref{ode3})), the main  difference lies in the range for $x$: in the euclidean case one has $\,x\in(0,1)$ while for the minkowskian case one has $\,x\in(1,+\nf)$. This explains why in \cite{ct} the authors could get also an explicit euclidean metric in despite of the fact that they were looking for minkowskian metrics.

The subsequent discussion is exactly the same as in section 5: one has three cases for the parameters 
and the B\"acklund transformations are the same but the sign fo $m'$ must be positive. 

An interesting question remains: what about the solution (\ref{Gromov})? It is of course still valid if 
we change $\,1-x\,\to \, x-1$ inside the logarithm. One is led to define
\[
U=\frac{(t+a)}{\tanh t}-1\qq a\in{\mb R}\qq\Longrightarrow\qq \ddot{U}-\frac 2{\sinh^2 t}\,U=0\]
which allows to write
\[
\frac{\mu}{2}=\frac{\dot{U}}{\dot{U}^2-U^2/\sinh^2 t}\qq\qq
\frac{\dot{\mu}}{4}=-\,\frac{U^2/\sinh^2 t}{(\dot{U}^2-U^2/\sinh^2 t)^2},\]
showing that the positivity constraint for $\,\dot{\mu}$ is not valid and so we must reject 
this solution. This case is meaningful only for the euclidean signature, in which case the metric is hyperk\"ahler, but does not give anything new for the minkowskian signature.

\subsection{Second case: $\,\mathbf{E=0}$}
We have
\beq
\ga\de=\frac 1{t}\qq\mbox{and}\qq \left(\frac{\ddot{\mu}}{2\dot{\mu}}+\frac 1t\right)^2+\dot{\mu}+\frac{2\mu}{t}+\frac{\mu^2}{t^2\,\dot{\mu}}=0.\eeq
This case is again special and does not reduce to Painlev\'e VI because it exhibits the Lie symmetry
\[\mu\to a\,\mu\qq t\to\frac ta\qq a\neq 0.\]
Defining the two invariants
\[u=t\,\mu\qq\qq v=t^2\,\dot{\mu}>0\]
brings the ode to the form
\beq
(u+v)^2\left[\left(\frac{dv}{du}\right)^2+4v\right]=0.\eeq
We get two solutions:
\brm
\item A singular one: $\ u+v=0$ which gives $\,\dst \mu=-\frac{\mu_0}{t}$ which 
is acceptable for $\mu_0>0$.
\item The general solution cannot be real due to the constraint $\,v>0$ and must be rejected.
\erm
The remaining functions, using (\ref{fctsmu}), are
\[\de^2=\frac{\ka}{\sqrt{\mu_0}}\qq\qq \be^2=\be_0^2\,t^{-(\mu_0+2)}.\]
Up to scalings of $\si_2$ and $\si_3$ and the definitions
\[\mu=\frac 2{2+\mu_0}\qq\qq \tau=t^{-\mu},\]
we get eventually Siklos metric \cite{Si} as written in \cite{ctbis}
\beq
g_S=-\mu^2\,d\tau^2+\tau^2\,\si_1^2+\frac{\mu}{2(1-\mu)}\tau^{2\mu}\,\si_3^2+
\left(\si_2+\tau^{\mu}\,\si_3\right)^2,\qq \mu\in\,]0,1[.\eeq 
This metric is rather special since it was proved in \cite{ctbis} that it 
descibes a pp-wave with a strong isometry enhancment up to six Killing vectors.

\subsection{Third case: $\,\mathbf{E<0}$}
We have 
\beq
\ga\de=\frac{\sqrt{|E|}}{\sin(\sqrt{|E|}t)}.\eeq
Scalings of $\mu$ and $t$ allow to take $\,E=-1$. The ode (\ref{sysODE}(b)) is now
\beq
\left(\frac{\ddot{\mu}}{2\dot{\mu}}+\cot t\right)^2+\dot{\mu}+2\cot t\,\mu+4
+\frac{\mu^2}{\sin^2 t\,\dot{\mu}}=0\eeq
and the metric
\beq
g=\be^2\left(\frac{dt^2}{\sin^2 t}+\si_1^2\right)+\frac{\sqrt{\dot{\mu}}}{\sin t}\,\si_3^2
+\frac 1{\sin t\,\sqrt{\dot{\mu}}}\,(\si_2+\mu\,\si_3)^2\eeq
with
\beq
\frac{\dot{(\be^2)}}{\be^2}=\frac{\ddot{\mu}}{2\dot{\mu}}+\mu-\cot t.\eeq
One can still define, but this is quite formal  
\[x=\frac 12(\cot t+1)\qq\qq \mu(t)=-2\,m(x)  \qq\qq m'=D_x\,m\]
transforming the previous equation into
\beq
x^2(x-1)^2(m'')^2=-4m'(xm'-m)^2+4m'^2(xm'-m)-4m'^2.\eeq
We get once more the same structure but for a change of sign in the last term. One can 
use again Okamoto results but for $\,b_1^2=-4$ and $\,b_4^2=-4$ since the parameters 
$\,(\alf,\,\be,\,\ga,\de)$ become complex. For $\,b_3^2=-4$ we get real parameters
\[b_1=b_4=0\qq\qq b_3=\pm 2i\qq\qq\alf=0\quad\be=2\quad\ga=-2\quad\de=0.\]
The B\"acklund transformation is
\beq
m=\frac 1{4y(y-1)(y-x)}\Big(x^2(x-1)^2(y')^2+4(y-x)^2\Big).\eeq 
The classical solutions for $y$ are now complex but they are again mapped into $\,m\equiv 0$. 
Furthermore very little seems to be known on Painlev\'e VI when Okamoto parameters become complex.

The derivative of $\,m$ is
\beq
m'=-\frac 1{4y(y-1)}\frac{\Big\{x^2(x-1)^2(y')^2+4(y-x)^2\Big\}}{(y-x)^2}\eeq
so there are are further restrictions on $\,y$ since $m'$ must be positive.

\subsection{A B\"acklund transform}
Since the minkowskian case is now completely discussed it is time to compare with the results 
of \cite{ct}. Some computational work shows that their metric is
\[g=\be^2\left(-\frac{dx^2}{x(x-1)}+\si_1^2\right)+\wti{\ga}^{\,2}\,\si_3^2
+\wti{\de}^{\,2}(\si_2+\wti{\mu}\,\si_3)^2\qq\quad \be^2=\frac{e^{u_1}}{4}.\]
As already observed for $x\in(1,+\nf)$ it is minkowskian while for $x\in (0,1)$ it is  
euclidean. Using their equation (2.8) one gets
\beq\label{CT1}
\frac d{dx}\,\ln(\be^2)=\frac{2x-1+y}{x(x-1)}\eeq
where $y$ is a solution of
\beq\label{CT}
x^2(x-1)^2\,(y'')^2=-4y'(xy'-y)^2+4(y')^2(xy'-y)+5\,(y')^2+4y'.\eeq 
The changes
\[x=\frac 12\,(\coth t+1)\qq y=-\frac{\wti{\mu}}{2}\]
transform (\ref{CT}) into
\beq\label{CT2}
\left(\frac{\ddot{\wti{\mu}}}{2\dot{\wti{\mu}}}+\coth t\right)^2+\dot{\wti{\mu}}+2\,\coth t\,\wti{\mu}-5+\frac{(\wti{\mu}^2-4)}{\sinh^2 t\,\dot{\wti{\mu}}}=0.
\eeq
Comparing then relation (\ref{CT1}) and (\ref{fctsmu}) we obtain the desired B\"acklund:
\beq\label{B1}
\wti{\mu}=\frac{\ddot{\mu}}{2\dot{\mu}}+\mu+\coth t.\eeq
Let us prove:

\begin{nth} The B\"acklund transformation (\ref{B1}) does transform (\ref{finEp}) into (\ref{CT2}).
\end{nth}

\nin{\bf Proof:} 
Equation (\ref{odebeta}) gives 
\beq\label{pr1}
\sinh^2 t\,\dot{\wti{\mu}}=\frac{\mu^2}{\dot{\mu}}.\eeq
Using this relation to get rid of $\dot{\mu}$ in (\ref{finEp}) gives a second degree equation 
for $\mu$, the solution of which is the inverse B\"acklund
\beq\label{Binv}
\mu=\frac{\sinh^2 t\,\dot{\wti{\mu}}}{1+\sinh^2 t\,\dot{\wti{\mu}}}
\Big(\wti{\mu}-\coth t\pm\sqrt{\De}\Big)\qq \De=-\dot{\wti{\mu}}
-2\coth t\,\wti{\mu}+5-\frac{(\wti{\mu}^2-4)}{\sinh^2 t\,\dot{\wti{\mu}}}.\eeq 
Differentiating (\ref{pr1}) and getting rid of $\mu$ using (\ref{Binv}) we end up with
\beq
\ddot{\wti{\mu}}+2\coth t\,\dot{\wti{\mu}}=\pm 2\,\dot{\wti{\mu}}\,\sqrt{\De}.\eeq
Squaring of both sides gives (\ref{CT2}) and ends the proof.$\quad\Box$

It follows that in \cite{ct} the case $\,E<0$ was missing. The reason for this was explained in the introduction: the authors presented a solution of the Einstein equations without deriving it from the basic field equations as it is customary.

\section{Integrable geodesic flows}
As is well known the geodesic flow, for a manifold equipped with some metric $g$, is generated by 
the quadratic hamiltonian
\beq
H=\frac 12\,g^{\mu\nu}\,P_{\mu}\,P_{\nu}\eeq
and the Killing vectors $\,K^{\mu}_i$ generate linear conserved quantities
\beq
K_i=K^{\mu}_i\,P_{\mu}\qq\qq\{H,K_i\}=0.\eeq
The integrability, in Liouville sense, of the geodesic flow hinges on the existence of 4 
algebraically independent quantities (one of them being the hamiltonian) which are in involution 
with respect to the Poisson bracket. In the metrics discussed in this work we always have 
two commuting Killing vectors which are suited to our aim, so we just need one extra conserved 
quantity in order to reach integrability.

There are two useful tensors which can help us: Killing-Yano (K-Y) and Killing-St\"ackel (K-S) tensors. 
A K-Y tensor is defined by
\beq
Y_{\mu\nu}=-Y_{\nu\mu}\qq\&\qq\nabla_{(\mu}\,Y_{\nu)\rho}=0,\eeq
while for a (quadratic) K-S tensor 
\beq
S_{\mu\nu}=S_{\nu\mu}\qq\&\qq\nabla_{(\mu}\,S_{\nu\rho)}=0.\eeq
A K-S tensor generates a quadratic conserved quantity
\beq
{\cal S}=S^{\mu\nu}\,P_{\mu}\,P_{\nu}\qq\qq\{H,{\cal S}\}=0,\eeq
while the symmetric product of two K-Y tensors (possibly a square) give a K-S tensor. 

An important observation is that Liouville integrability ensures the existence of separating coordinates 
for the Hamilton-Jacobi equation. This helps to make life easier since Levi-Civita has given the 
necessary conditions for the separability of Hamilton-Jacobi, see for instance \cite{Pe}. 
For a metric of the form (\ref{classemet}) the Levi-Civita constraints are merely:
\beq\label{LC}
\ga^2=C\,\be^2\qq\qq \dot{\mu}(t)=0.\eeq
By scalings one can reduce the constant $\,C$ to be one, and the second constraint holds only 
for the metrics with $\ka=0$. So we have the short list
\brm
\item The Kinnersley metric (\ref{Ki}),
\item The ``little" hyperk\"ahler metric (\ref{kazero}) for $E=0$,
\item The two euclidean metrics (\ref{kazero}) for $\,E\neq 0$.
\erm
To achieve integrability we need to exhibit a quadratic Killing tensor for these metrics. In 
agreement with the general results \cite{wp}, \cite{Co} for Petrov type D metrics we will check  
that they do exhibit at the same time a K-Y and a K-S tensor. 

\subsection{Kinnersley metric}
The Killing vectors produce 4 conserved quantities
\beq\left\{\barr{l}
K_1=-X\,P_X+z\,P_z\qq K_2=P_y\qq K_3=P_z\\[4mm]\dst 
K_4=-zX\,P_X+\frac{2l}{X}\,P_y+\frac 12\left(z^2-\frac 1{X^2}\right)P_z\earr\right.\qq X=e^{-x}.
\eeq
This metric exhibits the Killing-Yano tensor 
\beq
{\cal Y}=Y_{\mu\nu}\,dx^{\mu}\wedge dx^{\nu}=l\,dt\wedge(\si_2+2l\,\si_3)-t(t^2+l^2)\,\si_3\wedge\si_1.\eeq
Its square gives the K-S tensor 
\[{\cal S}_0=S_0^{\mu\nu}\,P_{\mu}\,P_{\nu}\qq\quad S_0^{\mu\nu}=Y^{\mu\si}\,g_{\si\tau}\,Y^{\tau\nu}\]
but it is fully reducible since we have
\[{\cal S}_0=l^2\,H-K_1^2-4l^2\,K_2^2+2\,K_3\,K_4.\] 
Besides there is another quadratic K-S tensor 
\beq
{\cal S}\equiv S^{\mu\nu}\,P_{\mu}\,P_{\nu}\eeq
which can be written
\beq 
{\cal S}=\frac{X^2}{\De}\,P_X^2-P_t^2+\frac 1{X^2\De}\,(P_z-2l\,X\,P_y)^2
-4\,\frac{(mt+l^2)}{\De}\left(H-\frac{t(t-m)}{\De}\,P_y^2\right)\eeq
with $\,\De=t^2-2mt-l^2$. We have checked the irreducibility of this K-S tensor and its 
invariance under all the isometries.

The Liouville integrability follows from the existence of the following 4 independent quantities
\[H\qq K_2\qq K_3 \qq {\cal S}\]
which are in involution for the Poisson bracket. As already mentioned the Hamilton-Jacobi equation 
does admit separation of variables, but since the metric is Ricci-flat quantum integrability is also 
preserved within ``minimal quantization" (see \cite{dv}) leading to the separability of the 
Schr\" odinger equation as well.

\subsection{The ``little" hyperk\"ahler metric}
The analysis of which Multi-Centre metrics have an integrable  geodesic flow, initiated in \cite{gr}, was completed in \cite{Va}. In this last reference the dipolar potential 
appears as the special case of  the first dipolar breaking of Taub-NUT with $m=v_0=0$ and 
${\cal F}=a$. The hamiltonian is
\beq
H=\frac V2\,P_z^2+\frac 1{2V}\left\{ \left(P_u+\frac{av}{r^3}P_z\right)^2
+\left(P_v-\frac{au}{r^3}P_z\right)^2+P_w^2\right\}.\eeq
The Killing vectors give the conserved quantities
\beq\left\{\barr{l} 
K_1=z\,P_z-u\,P_u-v\,P_v-w\,P_w\qq K_2=-v\,P_u+u\,P_v\qq K_3=P_z\\[4mm]\dst 
K_4=-z(u\,P_u+v\,P_v+w\,P_w)+\frac 12\left(z^2-\frac{a^2}{r^2}\right)P_z+
\frac ar\,(u\,P_v-v\,P_u).\earr\right.\eeq
Among the angular momentum components
\beq
\vec{L}=\vec{r}\wedge\vec{P}\qq\qq \vec{r}=(u,v,w)\eeq
only $L_3$ is conserved, since it is an isometry. From the general structure of the K-S tensor given 
in \cite{Va} some work is needed to extract the special case $m=v_0=0$ and to express it in terms of 
the true momenta $\,P_{\mu}$. One gets
\beq
{\cal S}\equiv S^{\mu\nu}\,P_{\mu}\,P_{\nu}=\vec{L}^{\,2}-2a\,\frac wr\,H\eeq
which is irreducible and invariant under the action of the four Killing vectors.

\subsection{Euclidean Kinnersley metrics}  
For the metric with $\,E>0$ in formula (\ref{kazero}), the isometries are the same as for their 
minkowskian partner. The K-Y tensor is now
\beq
{\cal Y}=Y_{\mu\nu}\,dx^{\mu}\wedge dx^{\nu}=l\,dt\wedge(\si_2+2l\,\si_3)+t(t^2-l^2)\,\si_3\wedge\si_1
\eeq
and its square is still reducible. The K-S tensor is 
\beq
{\cal S}=-\frac{X^2}{\De}\,P_X^2+P_t^2-\frac 1{X^2\De}(P_z-2l\,X\,P_y)^2
+4\frac{(mt-l^2)}{\De}\left(H+\frac{t(t-m)}{\De}\,P_y^2\right).\eeq
with $\,\De=t^2-2mt+l^2$. 

It is also valid for the metric with $\,E<0$ since we have $\,g_-=-g_+$.

\subsection{A remark on Siklos metric}
From Levi-Civita conditions (\ref{LC}) we know that the geodesic flow cannot be integrable for this metric. However, in a recent work \cite{kt} it has been shown how to construct a K-S tensor 
for the pp-waves. Now it happens that all the necessary tools: the Killing vectors and an 
homothetic Killing vector were given in \cite{ctbis}. It is convenient to use null coordinates 
for which Siklos metric can be written
\beq 
g_S=-2\mu\,du\,dv+\frac{(2-\mu)}{2(1-\mu)}\,dy^2+2u^{\mu}\,dy\,dz+u^{2\mu}\,dz^2\eeq
giving for hamiltonian
\beq
\mu\,H=-P_u\,P_v+2(1-\mu)\,P_y^2-2(1-\mu)\,u^{-\mu}\,P_y\,P_z+(2-\mu)\,u^{-2\mu}\,P_z^2.\eeq
The homothetic Killing vector and the first 5 Killing vectors are
\beq\left\{\barr{l}
Y^H=u\,P_u+v\,P_v+y\,P_y+(1-\mu)\,P_z,\qq K_1=uP_u-vP_v-\mu\,zP_z,\\[4mm]
K_2=P_y\qq K_3=P_z \qq K_4=P_v\qq K_5=yP_v+2(1-\mu)u\,P_y-2u^{1-\mu}\,P_z\earr\right.
\eeq
whereas the 6th one has a very special form \cite{ctbis} for $\mu=1/2$. One has
\beq\left\{\barr{ll}
\mu\neq 1/2\qq & \dst K_6=z\,P_v-2u^{1-\mu}\,P_y-\frac{(2-\mu)}{(2\mu-1)}\,u^{1-2\mu}\,P_z\\[4mm]
\mu = 1/2\qq & \dst K_6=z\,P_v-2\sqrt{u}\,P_y+\frac 32\,\ln(u)\,P_z.\earr\right.\eeq
Keane and Tupper have shown that the tensor
\[S_{\mu\nu}=k_{(\mu}\,Y^H_{\nu)}+\mu\,u\,g_{\mu\nu}\]
is indeed K-S. However one can check that for $\mu\neq 1/2$  the corresponding conserved 
quadratic quantity is fully reducible
\[S^{\mu\nu}\,P_{\mu}\,P_{\nu}=-\,K_1\,K_4+K_2\,K_5+(1-2\mu)\,K_3\,K_6\]
as well as for  $\mu=1/2$ 
\[S^{\mu\nu}\,P_{\mu}\,P_{\nu}=-\,K_1\,K_4+K_2\,K_5+\frac 32\,K_3^2.\]
We conclude to the full reducibility of this K-S tensor in all cases, in agreement with the Levi-Civita conditions. 

\section{No hyperk\"ahler metrics except for type III}
We have seen that for Bianchi type III non-diagonal hyperk\"ahler metrics do exist. This is very special to type III since one has the following:

\vspace{3mm}
\begin{nth} Among Bianchi B empty space metrics (excluding type III) there are no hyperk\"ahler metrics.\end{nth}

\nin{\bf Proof:}
For the ease of the reader we will split up the proof into three steps.

\vspace{3mm}
\nin{\em Step 1: the self-dual spin connection}

\nin Let us write the metric as
\beq
g=dt^2+\ga_{ij}(t)\,\si_i\,\si_j\qq\qq i,j=1,2,3\eeq
where the matrix $\ga$ can be taken symmetric  and must be positive definite and invertible. 
The Maurer-Cartan invariant 1-forms have for differentials
\beq
d\si_i=\frac 12\,C_{i;st}\,\si_s\wedge\si_t\eeq
or explicitly for the type B Lie algebras
\beq
d\si_1=0\qq d\si_2=n_2\,\si_3\wedge\si_1+a\,\si_1\wedge\si_2\qq 
d\si_3=-a\,\si_3\wedge\si_1+n_3\,\si_1\wedge\si_2\eeq
with the table
\beq\label{tableB}
\barr{cccrl} &\qq a &\qq n_2 &\qq n_3 &\qq\\[4mm] {\rm type}\ III &\qq 1 &\qq 1 &\qq -1 & \\ 
{\rm type}\ IV &\qq 1 &\qq 0 &\qq 1 & \\{\rm type}\ V &\qq 1 &\qq 0 &\qq 0 & \\ 
{\rm type}\ VI_h &\qq  \sqrt{-h} &\qq 1 &\qq -1& \quad (h<0,\  h\neq -1)\\ 
{\rm type}\ VII_h &\qq \sqrt{h} &\qq 1 &\qq -1 & \quad (h>0)\earr\eeq
It is well known that one can find some invertible symmetric matrix $\rho$ such that 
$\,\ga=\rho^{\,t}\,\rho$. So it is convenient to define the tetrad
\beq
e_0=dt\qq\qq e_i=\rho_{ij}(t)\,\si_j\qq i,j=1,2,3,\eeq
and some algebra gives for the self-dual components of the spin connection
\beq
\om^+_i\equiv \om_{0i}+\frac 12\,\eps_{ijk}\,\om_{jk}=g_i(t)\,dt+m_{ij}(t)\,e_j,\eeq
where
\beq\label{sdOm}
m_{ij}=L_{ij}-(\dot{\rho}\,r)_{(ij)}\qq\qq g_i=-\eps_{ijk}\,(\dot{\rho}\,r)_{[jk]}\qq r=\rho^{-1}\eeq
and
\beq
L_{ij}=\frac 1{2\,\det\rho}
\Big({\rm Tr}(\rho\mu\rho)\,\de_{ij}-2(\rho\mu\rho)_{ij}\Big),\qq 
\mu_{ij}=\frac 12\,\eps_{ist}\,C_{j;st}\ \to\   
\left(\barr{rrr} 0 & 0 & 0\\ 0 & n_2 & -a\\ 0 & a & n_3\earr\right).\eeq
Let us notice that the matrix $\,L$ is {\em never} diagonal. Its skew-symmetric part is
\beq
L_{[ij]}=a\eps_{ijk}\,r_{1\,k}\qq\quad r=\rho^{-1}.\eeq
So if we impose $\om^+_i=0$ we must have $\,L_{[ij]}=0$ and since for all Bianchi type B 
Lie algebras $a\neq 0$ , this implies that $r_{1k}=0$ for $k=1,2,3$ contradicting 
the hypothesis that $\,\rho$ is invertible. So there are no hyperk\"ahler Bianchi B metrics with 
vanishing self-dual connection.

\vspace{3mm}
\nin{\em Step 2: the self-dual curvature}

\nin Recalling the triplet of 2-forms with self-dual curvature 
\[R_i^+=d\om^+_i-\frac 12\,\eps_{ijk}\,\om^+_j\wedge\om^+_k\]
an easy computation shows that
\beq\label{sdR}
R^+_i=0\qq\Longleftrightarrow\qq \left\{
\barr{ll} \dot{M}_{ij}=\eps_{ist}\,g_s\,M_{tj}\qq & (a)\\[4mm]
M_{ij}\,C_{j;st}=\eps_{ijk}\,M_{js}\,M_{kt}\qq & (b)\earr\right.\qq M=m\,\rho.\eeq
Differentiating (\ref{sdR})(b) and using (\ref{sdR})(a) gives nothing but an identity.

\vspace{3mm}
\nin{\em Step 3: structure of the matrix $M$ and conclusion}

\nin Writing the matrix $M$ in terms of its column vectors
\[M=\Big(\,M_1\,M_2\,M_3\Big)\]
the relations (\ref{sdR})(b) split up into the system
\beq\left\{\barr{l}
A\,M_2=a\,M_2+n_3\,M_3\\[4mm] A\,M_3=-n_2\,M_2+a\,M_3\\[4mm] 0=\eps_{ijk}\,M_{j2}\,M_{k3}\earr\right.
\qq A_{ij}=\eps_{isj}\,M_{s1}.\eeq
Using the table (\ref{tableB}) and observing that the skew-symmetric matrix $A$ may have 
only 0 as a real eigenvalue, it is easy to prove that we must have $M_2=M_3=0$. This fails to be 
true for the type III.

Denoting by $b_1,\,b_2,\,b_3$ the components of the column vector $M_1$, we can compute the 
skew-symmetric part of the matrix $\,m=M\,r$ and by the first relation in (\ref{sdOm}) we must have  $\,m_{[ij]}=L_{[ij]}$. This gives a linear system
\beq\barr{r}
b_2\,r_{11}-b_1\,r_{12}+2a\,r_{13}=0\\[4mm]
2a\,r_{11}+b_3\,r_{12}-b_2\,r_{13}=0\\[4mm]
-b_3\,r_{11}+2a\,r_{12}+b_1\,r_{13}=0\earr\eeq
the determinant of which is $\,2a(b_1^2+b_2^2+b_3^2+4a^2)\neq 0$. This implies that 
$r_{11},\,r_{12}$ and $r_{13}$ do vanish, contradicting the hypothesis that $\rho$ must be invertible.
 
\nin This ends up the proof.$\quad\Box$

This Proposition implies that there are not even {\em flat} euclidean metrics, since 
these would be hyperk\"ahler.

\section{Conclusion}
We hope to have clarified and simplified some aspects of the analysis developed in \cite{ctbis} and 
\cite{ct}. In our opinion the most striking discovery of Christodoulakis and Terzis is that, for most of the Bianchi B metrics, the (empty space) Einstein equations do exhibit the Painlev\'e property no matter what the signature be.  Such a phenomenon was observed earlier for the diagonal metrics of types VI$_0$ and VII$_0$: Lorenz-Petzold \cite{LP} reduced them to Painlev\'e III. This shows that the theoretical prejudice according to which this should happen rather for euclidean metrics with self-dual Weyl tensor is far from being the whole story. 

Certainly nobody would bet that the empty space 
Einstein equations do enjoy in general the Painlev\'e property, however this may happen for 
particular classes of metrics! So it would be of great interest to have an answer to the 
question: which classes of metrics do exhibit the Painlev\'e property? In view of the formidable difficulty of this question its answer is probably ``the stuff the dreams are made of".


\end{document}